\documentstyle[preprint,aps]{revtex}

\begin{document}

\gdef\journal#1, #2, #3, 1#4#5#6{{#1~}{#2} (1#4#5#6) #3}
\gdef\ibid#1, #2, 1#3#4#5{{\bf #1} (1#3#4#5) #2}


\title{Quantum liquids of particles with
generalized statistics}

\author{Serguei B. Isakov}
\address{Department of Physics,  University of Oslo,
P.O. Box 1048 Blindern, N-0316 Oslo, Norway \\
and  Division de Physique Th\'eorique,
Institut de Physique Nucl\'eaire, 
Orsay Fr-91406, France\footnote{
Present address. 
 isakov@ipno.in2p3.fr
}
} 

\tighten
\maketitle
\begin{abstract} 

We  propose  
a phenomenological approach 
to quantum liquids of particles obeying generalized 
statistics of a fermionic  type,    
in the spirit of the Landau Fermi liquid theory. 
The approach is developed for fractional exclusion statistics.
We discuss both equilibrium 
(specific heat, compressibility, and Pauli spin susceptibility) 
and nonequilibrium (current and thermal conductivities, thermopower) 
properties.
Low temperature quantities have the same temperature 
dependences  as for the Fermi liquid, with the 
coefficients depending on the statistics parameter. 
The novel quantum liquids  provide explicit realization of 
systems with a non-Fermi liquid Lorentz ratio in two and more dimensions. 
Consistency of the theory is verified by deriving the 
compressibility  and $f$-sum rules.

\end{abstract}

\vspace{2cm}



\baselineskip=17pt

\narrowtext

\newpage

\section
{Introduction}
\label{introduction}

There has been increasing interest in generalized 
statistics for identical particles due to possible  
applications to fractional quantum Hall effect and  high-$T_c$ 
superconductivity \cite{anyons,FQHE-book}. 
Fractional exclusion statistics  \cite{Haldane91}, 
for which a single-state distribution function can be defined in the 
thermodynamic limit  \cite{I-MPLB,Wu94,Raj},  
has appeared in a system of  anyons in the lowest Landau level \cite{dVO94}, 
in Heisenberg \cite{LM} and Schr{\"o}dinger \cite{Poly89} 
quantization of identical particles (as fractional statistics in one 
dimension), 
in Calogero-Sutherland models \cite{ES-integrable}
and received much attention because of its relevance to 
the fractional quantum Hall effect, both to the bulk \cite{ES-FQH-bulk} and 
edge \cite{ES-FQH-edge} physics.     

It has been realized that the idea of a generalized 
exclusion principle \cite{Haldane91}  
is flexible enough to be able to accommodate    
some other statistical distributions \cite{OtherDistr},  
for instance  those for Gentile parafermions \cite{Gentile}.
All these distributions  belong to  
a class---we refer to the relevant statistics as  
{\em local}---for which  the average number of 
particles in a given quantum state $i$ 
 depends only on the Gibbs factor 
 $x_i = e^{(\mu-{\varepsilon}_i)/{k_B T}}$ for the same quantum state,  
$n_i = n(x_i)$. The function $n(x_i)$ is considered to be quite arbitrary 
up to natural physical constraints such as the existence of 
the Boltzmann limit for 
any statistical distribution for an ideal quantum gas  
when the mean occupation numbers  
are small,  $n_i\to x_i$ as $x_i\to 0$  \cite{I-IJTP}. 
If one specifies  the  equation for the distribution function to be 
\begin{equation}
\frac{n_i}{{\cal F}(n_i)}=e^{(\mu-{\varepsilon}_i)/{k_B T}} \, ,
\label{generic-distribution}\end{equation}
then the Boltzmann limit condition
implies  $ {\cal F}(n) \to 1 $ as $n\to 0$.

Low temperature thermodynamic quantities were calculated  for 
ideal gases of particles obeying exclusion and other local statistics 
\cite{ES-TD,IAMP,PhysicaA}. 
To discuss 
possibilities to detect generalized statistics in experiments, 
one normally has to have a theory of interacting particles. 
In this paper we suggest a phenomenological way to treat 
interaction  in systems of particles obeying local  statistics,  
in the spirit of Landau's approach to the Fermi liquid \cite{Landau}
(see also \cite{PN}).   
We restrict our attention to statistics of a  {\em fermionic  type}, 
whose zero temperature distribution functions are  step functions 
(with the height of the step generally different from unity).

In  Sect.~\ref{spinless} we formulate the notion of 
quantum liquids of particles obeying generalized  statistics  
and discuss low temperature equilibrium properties for 
spinless particles.   
Explicit formulas are  worked out for exclusion statistics. 
In Sect.~\ref{gas} we  study non-equilibrium quantities in  
a simpler case of noninteracting exclusion statistics 
particles in the relaxation time approximation. 
We show that the Wiedemann-Franz law holds, with a non-Fermi Lorentz ratio 
 depending on the statistics parameter, and  
discuss relation to the calculation of the Lorentz ratio for Luttinger 
liquids \cite{KaneFisher96}.
In Sect.~\ref{liquid}
we introduce the 
transport (kinetic) equation for the liquid  
and show that the Lorentz ratio 
is not modified by the interaction. 
To verify the theory, we check in Sect.~\ref{linearResponse} 
that the sum rules for the density response function are respected.
In Sect. \ref{spinning}
we discuss how spin of the particles can  be included. 
We conclude in Sect.~\ref{conclusion} with remarks on 
generic local statistics, 
non-linear dispersion of quasiparticles and  
possible applications to compressible 
quantum Hall states. 

\section{Liquid of spinless particles}
\label{spinless}
 
For an ideal gas of particles obeying  
fermionic  local statistics, by definition, the  zero temperature 
distribution function is constant 
(not necessarily equal to unity) for $p<p_0$ 
(inside a pseudo-Fermi sphere) and vanishes for $p>p_0$. 
In generalizing Landau's approach \cite{Landau} to such statistics, 
we assume that when interaction between particles is adiabatically 
switched on, there exists a liquid state which can be described,
at low temperatures, in terms 
of elementary excitations, quasiparticles (``dressed'' particles),  
which inherit quantum statistical properties of particles of the ideal gas. 
  
The latter means 
 that the entropy of the liquid, in terms of the 
quasiparticle distribution function $n_{\bbox{p}}$ (we consider 
spinless particles),  
is given by the same combinatorial expression as for a corresponding   gas. 
For exclusion statistics 
this implies  \cite{I-MPLB,Wu94}     
\begin{eqnarray}
S&=&\sum_{\bbox{p}} \Bigl\{ 
[1+(1-g) n_{\bbox{p}}] \,\ln[1+(1-g) n_{\bbox{p}}] \nonumber \\
&&-(1-g n_{\bbox{p}}) \,\ln (1-g n_{\bbox{p}})
- n_{\bbox{p}} \,\ln n_{\bbox{p}}
 \Bigr \} \; .
\end{eqnarray}
The statistics parameter $g$ is assumed  to be positive 
to avoid the Bose limit; $g=1$ corresponds to  Fermi statistics. 

The energy of the system is a functional of 
the  distribution of quasiparticles, $E=E[n_{\bbox{p}}]$.  
Variation of the energy 
\begin{equation}
\delta E = \sum_{\bbox{p}} {\varepsilon}_{\bbox{p}} \, \delta n_{\bbox{p}}
+ \sum_{\bbox{pp}'} 
f_{\bbox{pp}'} \,\delta n_{\bbox{p}} \,\delta n_{\bbox{p}'}
\label{deltaE}\end{equation}
defines the energy of a quasiparticle  ${\varepsilon}_{\bbox{p}}$ 
(which is a functional of the distribution function, 
${\varepsilon}_{\bbox{p}}={\varepsilon}_{\bbox{p}}[n_{\bbox{p}}]$) 
as well as  the quasiparticle interaction functional $f_{\bbox{pp}'}=
f_{\bbox{pp}'}[n_{\bbox{p}}] $.

Varying the entropy 
at constant total energy and  constant particle number, with the use of 
(\ref{deltaE}), yields an equation for the distribution function
\begin{equation}
\frac{ n_{\bbox{p}}}{(1-g n_{\bbox{p}})^g[1+(1-g) n_{\bbox{p}}]^{1-g}}=
e^{(\mu- {\varepsilon}_{\bbox{p}}[ n_{\bbox{p}}])/k_B T }  \;.
\label{basic-ES}\end{equation}

At zero temperature 
the distribution function is the same as for an ideal gas,  
\begin{equation}
 n_{\bbox{p}}= \Bigl\{  
\begin{array}{cl}
1/g  &  \quad |\bbox{p}|< p_{0} \\
0  &  \quad  |\bbox{p}|> p_{0}
\end{array} \; .
\label{n(T=0)}\end{equation}
This can be justified with the use of Eq.~(\ref{basic-ES}) 
and the zero temperature  dispersion of 
quasiparticles (\ref{dispersion}). 
 
The number of quasiparticles is equal 
to the number of particles 
and hence the relation between the pseudo Fermi momentum 
and the density is the same as for an  ideal gas,  
\begin{equation}
N= \int_{|\bbox{p}|< p_0} \frac{ d^d \bbox{p}\, V}{(2\pi \hbar)^{d}}  
\frac{1}{g} \; . 
\label{N}\end{equation}
With   
$d^d \bbox{p} = p^{d-1}\, dp\, d{\Omega}_{{\bbox p}}$, where 
$ d{\Omega}_{{\bbox p}}$ stands for  integration over angles of ${\bbox p}$ 
so that $\int  d{\Omega}_{{\bbox p}}=S_{d-1}$, and 
$S_d = 2\pi^{d/2}/\Gamma ({ \textstyle \frac12}d )$ is the 
surface of a $d$-dimensional sphere of unit radius ($S_0\equiv 2$), 
Eq.~(\ref{N}) implies a relation between the pseudo Fermi momentum and the 
particle number density $\rho=N/V$  
\begin{equation}
\left(  \frac{p_0}{2\pi\hbar} \right)^d =g \,\frac{ \rho \, d}{S_{d-1}} \; .
\label{N-p0}\end{equation}
Note that in microscopic theory, Eq.~(\ref{N-p0}) has be proved,  
as a generalization of the Luttinger theorem \cite{LuttingerTh}.

For $T=0$ the quasiparticle energy functional becomes a function 
since the distribution function is fixed.  
We assume the  standard linear dispersion for quasiparticles 
near the pseudo Fermi energy
\begin{equation}
{\varepsilon}_{\bbox{p}} = \mu +v_0 (p-p_0) + \ldots 
\label{dispersion}\end{equation}
where $v_0 \equiv {p_0}/{m^*}$, and $m^* $ is an effective mass.
The quasiparticle interaction functional also becomes a function. 
We will only need to know the 
quasiparticle interaction function on the pseudo  Fermi surface,
$f_{{{\bbox p}}_0 {{\bbox p}}'_0}$. 
For a spherically symmetric case, it  
depends only on 
the angle $\theta$ between $\bbox{p}$ and  $\bbox{p}'$: 
$f_{{{\bbox p}}_0 {{\bbox p}}'_0}=f(\theta)$,  
with the function $f(\theta)$ even, 
due to the symmetry $f_{{\bbox p} {\bbox p}'}=f_{{\bbox p}' {\bbox p}}$.  

A relation between the effective mass and the quasiparticle interaction
function follows from the Galilean invariance. The latter
implies 
$\sum_{{{\bbox p}}} {{\bbox p}} n_{{\bbox p}} 
=\sum_{{\bbox p}}  m ({{\partial}{\varepsilon}_{{\bbox p}}[ n_{{\bbox p}}] }/
{{\partial}{\bbox p}}) n_{{\bbox p}} $ or, upon varying 
w.r.t.  $n_{{\bbox p}}$ \cite{Landau} 
\begin{equation}
\frac{{\bbox p}}{m} =\frac{{\partial}{\varepsilon}_{{\bbox p}}}{{\partial}
{\bbox p}} 
-\sum_{{{\bbox p}}'} f({\bbox p}, {{\bbox p}}')  \frac{{\partial}
n_{{{\bbox p}}'}}{{\partial}{{\bbox p}}' } \;. 
\label{Galilean}\end{equation}
Multiplying (\ref{Galilean}) by ${\bbox p}$, choosing  ${\bbox p}$ to belong 
to the pseudo 
Fermi surface  and using 
the zero temperature distribution function 
with 
${{\partial}n_{{{\bbox p}}'}}/{{\partial}{{\bbox p}}' }=-({{\bbox p}}'/p')
\frac{1}{g} \delta (p'-p_0)$,
we obtain     
\begin{equation}
\frac{m^*}{m}=1+\frac{1}{g}\, \nu ({\varepsilon}_0)\,  
\overline{f(\theta)\,\cos \theta} \;.
\label{mass}\end{equation}
The bar denotes the average over angles in spherical coordinates in a 
$d$-dimensional  ${{\bbox p}}'$-space, with the polar axis 
along ${\bbox p}$ and polar angle $\theta$,  
$ \overline{A(\theta)} \equiv \int A(\theta) \,
( d\Omega_{{{\bbox p}}'} /S_{d-1}) $. 
We have also introduced the 
density of states on the pseudo Fermi surface
\begin{equation}
\nu({\varepsilon}_0) =\frac{ S_{d-1}}{(2\pi\hbar)^d}\, m^* \,p_0^{d-2}\, V\;.  
\label{nu}\end{equation}

One can introduce a dimensionless quasiparticle interaction function  
$ F(\theta)=  \nu ({\varepsilon}_0)  f(\theta), $
and  define Landau parameters $F_0$ and  $F_1$ by 
\begin{equation}
 \overline{F(\theta)}=F_0\,, \quad 
 \overline{F(\theta)\,\cos \theta}= \frac{1}{d} \, F_1 \;. 
\end{equation}
In one dimension, where the polar angle takes two values, $\theta =0,\pi$, 
 $F_0$ and $F_1$ are given by    
$ F_0= {\textstyle  \frac12 } [F(0) +F (\pi)]$ and $F_1= 
{\textstyle  \frac12 } [F(0) -F (\pi)] $. 
In two dimensions they are coefficients of an expansion  
$F(\theta) =\sum_{l=0}^{\infty} F_l \cos (l\theta)$, 
and in three dimensions they correspond to the 
standard expansion in Legendre polynomials, 
$F(\theta) =\sum_{l=0}^{\infty} F_l  P_{l}(\cos \theta)$. 

In the above  notation, (\ref{mass}) reads 
\begin{equation}
\frac{m^*}{m}=1+\frac{1}{g} \frac{F_1}{d} \; .
\label{m*}\end{equation}  

For calculating the specific heat one can use the dispersion law 
(\ref{dispersion}) 
for zero temperature. Finite temperature corrections to the dispersion
(from the second term in the r.h.s. of (\ref{deltaE}))
give  terms of higher order in the temperature and can be neglected.
The specific heat is thus obtained by the change $m\to m^*$ in the expression 
for an ideal gas given in Ref.~\cite{IAMP},  
\begin{equation}
{C}_V= \frac{\pi^2}{3}\nu({\varepsilon}_0)\, k_B^2  T     \; . 
\label{C}\end{equation}
To calculate 
the compressibility $\kappa$ at zero temperature,  
defined as $1/\kappa =N\rho ({\partial}\mu /{\partial}N)_{V}$, 
we note  that in varying $N$  there are two sources for 
changing the pseudo-Fermi energy,   
${\varepsilon}_{{\bbox p}} [n_{{\bbox p}}|_{T=0}]|_{p=p_0}$, due to the 
changes of   
 the distribution of quasiparticles and the pseudo-Fermi momentum. 
In this way we obtain 
${1}/{\kappa}= N\rho  (g+{F_0})/{\nu({\varepsilon}_0)}$, 
or in terms of  the first sound velocity,  
\begin{equation}
v_s^2=\frac{p_0^2}{d\,mm^*}\left(1+\frac{F_0}{g}  \right) .       
\label{v_s}\end{equation}
($v_s=p_0/m\sqrt{d}$ in the limit of vanishing interaction).

\section{ Transport properties in the relaxation time approximation}
\label{gas}
Four transport coefficients  relate electrical and 
thermal currents to changes in electrochemical potential  
and the temperature (we consider an isotropic liquid)  \cite{mermin} :
\begin{eqnarray}
\bbox{j} &= {\cal L}^{11} \left( {\bbox E}-\frac{\nabla \mu}{e} \right)  + 
 {\cal L}^{12}\, (- \nabla T)  \; , \nonumber \\
\bbox{j}^Q &= {\cal L}^{21} \left( {\bbox E}-\frac{\nabla \mu}{e} \right)  +   
{\cal L }^{22}\, (- {\nabla} T) \; .
\label{L}
\end{eqnarray}

We start with discussion of a simpler case of the  
transport equation for noninteracting 
exclusion statistics particles 
in the relaxation time approximation
\begin{equation}
\frac{{\partial}n}{{\partial}t} + {\bbox v} ({\varepsilon}) 
\frac{{\partial}n}{{\partial}{\bbox{r}}} +
e \bbox{E} \frac{{\partial}n}{{\partial}{\bbox p}} = - \frac{n-n^0}{\tau 
({\varepsilon})}
\label{tau}\end{equation}
allowing for dependence of the relaxation time 
$\tau ({\varepsilon})$ on the energy. 
Here $n^0$ is an equilibrium (uniform) distribution.
We do not specify the dispersion law of particles,  
only  assuming that  it is isotropic, ${\varepsilon}_{{\bbox p}}=
{\varepsilon}(p)$; 
${\bbox v} ({\varepsilon})= {\partial}{\varepsilon}/{\partial}\bbox{p} $ 
is the particle  
velocity.   
The carrier charge $e$ may be different from the charge of an electron 
and may have either sign. 

Consider the  distribution function with slowly varying chemical 
potential and temperature  
$
n=n\left( e^{[\mu (\bbox{r})  -{\varepsilon}_{\bbox{p}}]/k_B T(\bbox{r})}
\right)
$.
Then for the stationary solution to first order in gradients of the 
electrochemical potential and temperature we obtain from (\ref{tau})
$$ 
n=n^0-\tau ({\varepsilon}) \left\{ 
 {\bbox v} \cdot ( e \bbox{E}- \nabla \mu )  +\frac{\mu-{\varepsilon}}{T} 
({\bbox v} \cdot \nabla T )  
  \right\} \frac{{\partial}n^0}{{\partial}{\varepsilon}}.
$$
With this distribution function the transport coefficients in (\ref{L}) 
read 
$
 {\cal L}^{11}={\cal L}_0$, $
{\cal L}^{12}= -\frac{1}{T} {\cal L}^{21}
=\frac{1}{eT}{\cal L}_1 $, and 
${\cal L}^{22}= \frac{1}{e^2 T} {\cal L}_2 $, 
where
\begin{equation}
{\cal L}_r =-\int_{0}^{\infty} 
d{\varepsilon}\,({\varepsilon}-\mu)^r \, \sigma_F ({\varepsilon}) \, 
\frac{{\partial}n^0}{{\partial}{\varepsilon}} 
\label{Lr}\end{equation}
and   
\begin{eqnarray}
\sigma_F ({\varepsilon}) &=&e^2 \tau ({\varepsilon}) \sum_{{\bbox p}} 
\delta ({\varepsilon}-{\varepsilon}_{{\bbox p}}) v^2 ({\varepsilon})  
\nonumber \\
&=& e^2 \tau ({\varepsilon}) \nu ({\varepsilon}) v^2 ({\varepsilon}) 
\label{sigmaF}\end{eqnarray}
($ \nu ({\varepsilon})$ is the density of states) is the energy-dependent 
conductivity for a Fermi gas with the Fermi energy ${\varepsilon}_0$. 
The symmetry between ${\cal L}^{12}$ and ${\cal L}^{21}$  
manifests  Onsager's principle.  

At low temperatures integrals of the form 
(\ref{Lr}), involving the distribution function for exclusion statistics, 
can be evaluated using the expansion   \cite{IAMP}
\begin{eqnarray}
\int_{0}^{\infty} d{\varepsilon}\, n_g({\varepsilon}) {\cal G}({\varepsilon}) 
&=& \frac{1}{g} \int_{0}^{\mu} 
 d{\varepsilon}\,{\cal G}({\varepsilon}) \nonumber \\ 
&& +\frac{\pi^2}{6} {\cal G}'(\mu) (k_B T)^2 + \cdots  
\label{approxIntegrals}\end{eqnarray}
valid for functions ${\cal G}({\varepsilon})$ which vary on the scale 
${\varepsilon}_0$.
In this way we obtain the leading terms of the transport coefficients 
\begin{eqnarray}
 {\cal L}^{11}&=&  \frac{1}{g}\sigma_F ({\varepsilon}_0) \,, \quad  
{\cal L}^{22} = \frac{\pi^2 k_B^2 T}{3e^2} \sigma_F ({\varepsilon}_0) \,, 
  \nonumber \\
{\cal L}^{12}&=&-\frac{1}{T}  {\cal L}^{12} =-\frac{\pi^2 k_B^2 T}{3e}
  \left. \left(  \frac{{\partial}\sigma_F ({\varepsilon})}{{\partial}
{\varepsilon}} \right) \right|_{{\varepsilon}={\varepsilon}_0} .
\label{Lfinal}
\end{eqnarray}

The low temperature conductivity  
$\sigma= {\cal L}^{11}=\frac{1}{g}\sigma_F ({\varepsilon}_0)$
is independent of the temperature.  
The thermal conductivity relates the thermal current 
to the temperature gradient $\bbox{j}^{Q}= -K \, \nabla T$ provided 
$\bbox{j}=0$, thus yielding 
$ K ={\cal L}^{22}-{\cal L}^{12}{\cal L}^{21}/{\cal L}^{11}$.
From (\ref{Lfinal}) we observe that 
the ratio of the thermal conductivity 
to the electrical conductivity 
is proportional to the temperature  thus respecting  
the Wiedemann-Franz law but with the Lorentz number 
\begin{equation}
L \equiv \frac{K}{\sigma T} =g\, \frac{\pi^2}{3}
\left( \frac{k_B}{e} \right)^2  =
g \,L_F \;,
\label{Ltau}\end{equation}
differed  by the factor $g$ from its value  $L_F$ for a Fermi liquid. 

This observation tells us that the difference of the Lorentz number 
from its value for the Fermi liquid may signal a kinematic effect 
(change of statistics). 

The thermopower $Q={\cal L}^{12}/{\cal L}^{11}$ is 
\begin{equation}
Q= -g \frac{\pi^2 k_B^2 T}{3e}  
 \left. 
 \left(  \frac{{\partial}\ln \sigma_F ({\varepsilon})}{{\partial}
{\varepsilon}} \right) \right|_{{\varepsilon}={\varepsilon}_0} .
\label{Q}\end{equation}
For a free gas, assuming that 
$\tau ({\varepsilon}) \propto {\varepsilon}^{\lambda} $, one obtains 
$\sigma_F ({\varepsilon}) \propto {\varepsilon}^{\lambda + D/2} $
 and (\ref{Q}) reduces to 
$Q= - g (\lambda + d/2) ({\pi^2 k_B}/{3e}) (k_B T/{\varepsilon}_0)$. 

\smallskip 
{\em Relation to a  Luttinger liquid.---}
A Luttinger liquid \cite{LL} 
is characterized by the velocities of charge and current 
excitations, $v_N$ and $v_J$, and the sound velocity $v_s$, related by  
$ v_s^2=v_N v_J$. 
As two independent parameters, one can choose
 $v_s$ and a  parameter ${\cal K}$ such that
$ v_J= {\cal K} v_s$ and $v_N = v_s/{\cal K} $.   

By comparison of  the low temperature specific heat for a (spinless) 
Luttinger liquid
$C/V= \pi^2 k_B T /3 \hbar v_s$  \cite{LL} 
with that of 
an ideal gas of exclusion statistics particles,   
(\ref{C}) for
$F_0=F_1=0$, we observe that the thermodynamic quantities 
for the two systems are identical if 
\begin{equation}
g=1/{\cal K}   
\label{g-K}\end{equation}
(see also an argument by bosonization  \cite{WuYu}). 

Our calculation of the Lorentz ratio show that the 
the correspondence between a 
Luttinger liquid and an ideal gas of exclusion statistics particles can be
extended to non-equilibrium properties. Namely, by  
comparison (\ref{Ltau})  with the result for  the Lorentz ratio 
for a Luttinger liquid 
$L=L_F/{\cal K}$ \cite{KaneFisher96}, 
we see that the two expressions coincide  
under the above identification (\ref{g-K}).
      
\section{ Transport equation for the  liquid}
\label{liquid}
Now we turn to the liquid. 
Consider distributions corresponding to  
small departures from the equilibrium quasiparticle distribution function, 
$n_{{\bbox p}}({\bbox r},t)=n_{{\bbox p}}^{0}({{\varepsilon}}_{{\bbox p}}^0) 
+\delta n_{{\bbox p}}({\bbox r},t)$,  
where the superscript  0  labels equilibrium quantities.  
For short range interactions the corresponding change in energy of 
a quasiparticle can be written in a local form 
(i.e. depending only on $\delta n_{{\bbox p}}({\bbox r},t)$ at the same point):
$
\delta {\varepsilon}_{\bbox{p}}({\bbox r},t)= \sum_{\bbox{p}'} 
f_{\bbox{pp}'} \,  \delta n_{\bbox{p}'}({\bbox r},t)
$ 
\cite{PN}.
If $n^0$ is expressed in terms of the actual energy of quasiparticles, 
${\varepsilon}={\varepsilon}^0 +\delta {\varepsilon}$, 
the perturbed distribution function reads 
\begin{equation}
n_{{\bbox p}}({\bbox r},t) =n_{{\bbox p}}^{0}({{\varepsilon}}_{{\bbox p}}) 
+\delta {\tilde n}_{{\bbox p}}({\bbox r},t) \;,   
\label{delta-n}\end{equation}
where 
$
\delta {\tilde n} =
\delta {n}- \delta \varepsilon \, ({\partial}n^0 /{\partial}{\varepsilon})$  
is the quantity (rather than $\delta n $) 
determining the charge and thermal currents, 
$ {\bbox j}= -e \sum_{{\bbox p}} {\bbox v} \, {\delta {\tilde n}} $ and 
${{\bbox j}}^Q = -e \sum_{{\bbox p}} ({\varepsilon}-\mu) {\bbox v} \, 
{\delta {\tilde n}} $. 
The transport equation 
\begin{equation}
\frac{{\partial}\,\delta n  }{{\partial}t} 
+\frac{{\partial}{\varepsilon}^0}{{\partial}{{\bbox p}}}\frac{{\partial}\,
\delta n  }{{\partial} \bbox{r}}
-  \frac{{\partial}\,\delta {\varepsilon}}{{\partial} \bbox{r}}
\frac{{\partial}n^0}{{\partial}{\bbox p}}=I(n)
\label{transport}\end{equation}
has the same form as for the Fermi liquid, except for the 
collision integral $I(n)$, where factors $(1-n)$  
accounting for the Pauli exclusion principle
should be modified.
  
The form of the collision integral for ideal gases with 
arbitrary {\em local} statistics for identical particles 
was discussed  by the author   \cite{I-IJTP}. 
 It was argued in Ref.~\cite{I-IJTP} that in constructing 
collision integrals one should replace  factors 
accounting for the Pauli exclusion principle as follows:
\begin{equation}
(1-n) \to {\cal F}(n) \, ,
\label{change}\end{equation}
where the function $ {\cal F}(n)$ arises in the equation for the 
equilibrium distribution function (\ref{generic-distribution})  

For fractional exclusion statistics the function $ {\cal F}(n)$ can be 
read off 
from Eq. (\ref{basic-ES}): $ {\cal F}(n)=(1-gn)^{g}[1+(1-g)n]^{1-g} $.
The latter form was also found in discussions of 
the Boltzmann equation for exclusion statistics 
\cite{ES-BoltzmannEq}.  
 
Using the change (\ref{change}) we can   
write down the collision integral for the case of 
the residual interaction due to an elastic  scattering by impurities: 
$$
I(n_{{\bbox p}})= \frac{2\pi}{\hbar }\sum_{{\bbox p}'} w_{{\bbox p} {\bbox p}'}
\delta ({\varepsilon}_{{\bbox p}} -{\varepsilon}_{{\bbox p}'}) 
\left[n_{{\bbox p}} {\cal F}(n_{{\bbox p}'}) - n_{{\bbox p}'} 
{\cal F}(n_{{\bbox p}})
\right].
$$
Here $\frac{2\pi}{\hbar }w_{{\bbox p} {\bbox p}'}
\delta ({\varepsilon}_{{\bbox p}} -{\varepsilon}_{{\bbox p}'})$ is the 
transition probability for the 
scattering of a quasiparticle from state ${\bbox p}$ to state ${\bbox p}'$.
We assumed that the impurity atoms are randomly distributed and form 
a dilute gas so that they scatter independently; 
the detailed balancing principle $ w_{{\bbox p} {\bbox p}'}= w_{{\bbox p}' 
{\bbox p}}$ was also used. 

Using (\ref{delta-n}) we linearize the collision integral to obtain        
\begin{equation}
I(\delta \tilde n )= \frac{2\pi}{\hbar }\sum_{{\bbox p}'} w_{{\bbox p} 
{\bbox p}'}
\delta ({\varepsilon}_{{\bbox p}} -{\varepsilon}_{{\bbox p}'}) 
\left[ \delta \tilde n_{{\bbox p}} - \delta \tilde n_{{\bbox p}'} 
\right] \;.
\label{I-lin}\end{equation}
The transport coefficients can be  calculated using the linearized 
transport equation  
which to first order in gradients of the temperature and chemical potential 
reads 
\begin{equation}
\left( e{\bbox E}+\frac{\mu-{\varepsilon}}{T} {\bbox \nabla} T  \right) \cdot 
{\bbox v}\frac{{\partial}n^0}{{\partial}{\varepsilon}} =
  I(\delta {\tilde n}) \;,
\end{equation}
where ${\bbox v} ={\partial}{\varepsilon}^0 / {\partial}{\bbox p} $. 

For an  elastic scattering (in particular, for collision integrals of 
the form (\ref{I-lin}) when $w_{{\bbox p}{\bbox p}'}$ depends only on the 
angle between 
${\bbox p}$ and ${\bbox p}'$),   
collisions move particles on the constant energy 
surface, and one can seek the solution of the transport equation 
in the form (cf. relevant discussion for a Fermi liquid \cite{L-X})   
$ {\delta {\tilde n}} =
- ({{\partial}n^0}/{{\partial}{\varepsilon}})
[ e{\bbox E}+({\mu-{\varepsilon}}) ({\bbox \nabla} T /T ) ] \cdot 
{\bbox l}({\bbox p})$, 
where ${\bbox l}({\bbox p})$ satisfies the equation 
$I({\bbox l})=-{\bbox v}$.
Note that 
the Landau  parameters do not enter explicitly (they are absorbed in 
${\delta {\tilde n}}$).
As a result, we obtain the same form of the transport coefficients 
${\cal L}^{ab}$ as in the previous section for an ideal gas,  
with the energy dependent conductivity  (\ref{sigmaF})  replaced by 
$ \sigma_F ({\varepsilon}) =e^2  \sum_{{\bbox p}} 
\delta ({\varepsilon}-{\varepsilon}_{{\bbox p}}) \,\, \bbox{l} \cdot 
\bbox{v} $.

Then calculating the low temperature expansion as 
in Sect.~\ref{gas} yields the result (\ref{Lfinal}),
with the above $\sigma_F ({\varepsilon})$,  
and leads  to the previously found Lorentz ratio (\ref{Ltau}).  

The fact that the interaction  does not change the Lorentz  number 
(for a residual conductivity due to scattering by impurities)
suggests that this number may be a good quantity to check in searching 
for generalized statistics in experiments.


\section{Linear response and sum rules}
\label{linearResponse}

To derive linear  response functions from the transport equation, a 
harmonic perturbation of the density by a 
 scalar potential $\varphi ({\bbox q}, \omega)$ is applied 
corresponding to an external force  
$\bbox{F}= -i{\bbox q} \varphi ({\bbox q}, \omega) e^{i({\bbox q} 
{\bbox{r}} -\omega t)} + {\rm c.\, c.}$
acting on a liquid  \cite{PN}. 
This modifies the transport equation (\ref{transport}), written for   
the Fourier components, to   
\begin{eqnarray}
&&(\omega-{\bbox q}{\bbox v} ) \delta n_{{\bbox p}} ({\bbox q}, \omega) 
+ {\bbox q}{\bbox v} \, \frac{{\partial}n^0}{{\partial}{\varepsilon}_{
{\bbox p}}}  
\sum_{{\bbox p}'}f_{{\bbox p}{\bbox p}'} \delta n_{{\bbox p}'} ({\bbox q}, 
\omega)   \nonumber \\
&& + \; {\bbox q}{\bbox v} \, \varphi ({\bbox q}, \omega) 
\frac{{\partial}n^0}{{\partial}{\varepsilon}_{{\bbox p}}} = 
i I(\delta n_{{\bbox p}} ({\bbox q}, \omega) ) \, .
\label{kineq-linresp}\end{eqnarray}
By solving this equation for small $\varphi ({\bbox q}, \omega)$, one can  
find the density response function 
$\chi ({\bbox q}, \omega)=\langle  \rho ({\bbox q}, \omega) \rangle /\varphi 
({\bbox q}, \omega)$,
where $\langle  \rho ({\bbox q}, \omega) \rangle  =
\sum_{{\bbox p}} \delta n_{{\bbox{p}}}({\bbox q}, \omega)$ is the (average) 
induced density 
fluctuation. 

Consider the 
 collisionless regime where one can neglect the collision integral. 
In the {\em static} limit $\omega \ll q v_0$ 
we find from (\ref{kineq-linresp}) 
$ \delta n_{{\bbox{p}}}({\bbox q}, 0) = {\varphi ({\bbox q}, 0)}({
{\partial}n^0}/{{\partial}{\varepsilon}_{{\bbox{p}}}}) /(1+F_0/g)$.
This leads to  the response function 
\begin{equation}
\chi ({\bbox q}, 0) =-\frac{\nu({\varepsilon}_0)}{g+F_0}  \; .
\label{compress-sumrule}\end{equation}
By comparison with the expression for the first sound velocity 
(\ref{v_s}), we see that  Eq.~(\ref{compress-sumrule}) 
is consistent with the compressibility sum rule 
$\lim_{{\bbox q} \to 0} \chi ({\bbox q}, 0) =-{N}/{mv_s^2}$ \cite{PN}.

We note that the transport equation for the liquid is itself valid 
only for perturbations $\delta n$ varying on a macroscopic 
scale implying that eq.~(\ref{kineq-linresp}) 
holds only for ${\bbox q}$ small enough. 
  
In the {\em quasihomogeneous} limit, 
$q v_0 \ll \omega$, in calculating $\delta n_{{\bbox{p}}}({\bbox q}, \omega)$ 
in (\ref{kineq-linresp}) 
one has to keep terms up to second order in $q v_0 /\omega$, 
resulting in  
$\chi ({{\bbox q}, \omega})=Nq^2/m\omega^2$, 
which manifests the $f$-sum rule  
as $\omega \to \infty$ \cite{PN}. 

In  the hydrodynamic regime, 
where $\omega$ is much less than the 
collision frequency and $q \ell \ll 1$, with $\ell$ the mean free path,  
collisions dominate. 
In this case, all the deviations from the equilibrium distributions 
are damped out by the collisions, except for those 
of the form (near the pseudo-Fermi surface)
$\delta n^{(0)}$ (constant) and $\delta n^{(1)} \cos \varphi $, where 
$\varphi$ is the angle between $\bbox{k}$ and $\bbox{p}$. 
The latter contribute into the particle number and momentum fluxes 
respectively and for them the collision integral vanishes 
due to the particle number and momentum conservation. 

Integrating Eq.~(\ref{kineq-linresp}) over momenta, 
with ${\bbox q}$ as the  polar axis, 
with the weights 1 and $\bbox{p}$ respectively, one obtains 
 two coupled equations for 
$\delta n^{(0)}$ and $\delta n^{(1)}$ solution of which results in  
$ \chi ({\bbox q}, \omega)= (N/m)/ (\omega^2 / q^2 - v_s^2)$
leading to the same static and homogeneous limits as in the 
collisionless case. 
We thus conclude that our phenomenological theory  
respects  the compressibility and 
$f$-sum rules both in the collisionless  and in the  hydrodynamic regimes.

\section{Liquid of spinning particles. Spin susceptibility}
\label{spinning}

Consider spinning particles, of spin $1/2$  for definiteness, 
with $\alpha= \frac12, -\frac12$ labeling the spin projections  
(we also use   $\alpha= \uparrow , \downarrow$ in subscripts). 

There is an ambiguity with assigning statistics to spinning 
particles.
In general, one can introduce a statistics matrix $g_{\alpha\alpha'}$
(four parameters) determining how particles with  different 
spin projections  affect each 
other. There are two special cases, 
$g_{\alpha\alpha'}=g \delta_{\alpha\alpha'} $ and  
$g_{\alpha\alpha'}=g  $
involving a single statistics parameter. 
In this paper we discuss the case 
$g_{\alpha\alpha'}=g \delta_{\alpha\alpha'} $   which means 
that particles of different spin projections are not connected 
statistically (like ordinary fermions which are recovered for $g=1$). 

Now the energy of a quasiparticle and the Landau
 interaction function acquire additional spin indices.
 All the formulas of the previous sections are valid upon replacing the
 quasiparticle interaction function by its symmetric combination
 $f(\theta) \to f_{s}(\theta)= \frac12 [f_{\uparrow \uparrow}(\theta)
 +f_{\uparrow \downarrow}(\theta)]$
 (and similarly for the associated Landau parameters) provided, in addition,
 $\nu({\varepsilon}_0)$ is as twice as given by (\ref{nu}), 
due to the two spin polarizations.
 The latter definition of $\nu({\varepsilon}_0)$ is assumed in the rest 
of this section. 

 In the presence of the magnetic field, the variation of the energy of a
 quasiparticle reads (assuming the gyromagnetic ratio two)
 \begin{equation}
 \delta {{\varepsilon}}_{\bbox{p}\alpha}= -2 \mu_{B} \alpha B 
 + \sum_{\bbox{p}'{\alpha}'}
f_{\bbox{p}\alpha,\bbox{p}'\alpha'} \,\delta n_{\bbox{p}'\alpha'} \;.
 \label{energy-B}\end{equation}
With  $\delta n_{\bbox{p}\alpha} =
({\partial}n_{\bbox{p}\alpha}/{\partial}{\varepsilon}_{\bbox{p}\alpha}) \,
\delta 
{\varepsilon}_{\bbox{p}\alpha}$,
the solution of (\ref{energy-B}) to first order in $B$  at zero temperature
yields modification of the magnetic energy due to the interaction 
$ \delta{{\varepsilon}}_{\bbox{p}\alpha}=-{2 \mu_{B} 
\alpha B}/({1+{F_{0}^{a}}/{g}}) $, 
where $F_{0}^{a}=\frac12 (F_{\uparrow \uparrow, 0}
 -F_{\uparrow \downarrow, 0})$ 
 is an antisymmetric combination of the Landau
  parameters. From the variation of the magnetization
  $\delta M = \sum_{\bbox{p}{\alpha}} 2 \mu_{B} \alpha 
  \delta n_{\bbox{p}\alpha}$
we find the magnetic susceptibility 
\begin{equation}
\chi =\frac{\mu_{B}^2}{V} \frac{\nu({\varepsilon}_0)}{g+{F_0^a}} \;.
\label{chi}\end{equation}
Without including liquid effects ($F_0^a =0$), this expression reduces
to the magnetic susceptibility of an ideal gas of exclusion statistics
particles \cite{PhysicaA} (which means   that the authors of
Ref.~\cite{PhysicaA} used implicitly the statistics matrix
of the form $g_{\alpha \alpha'}=g \delta_{\alpha \alpha'} $).
Note that the Wilson ratio
$R_{W}=(\pi^2 k_{B}^2 T \chi V / 3\mu_B^2 C ) $,
 which is equal to $R_{W}= 1/(g+F_{0}^{a})$ 
according to (\ref{C}) and (\ref{chi}), differs 
from its Fermi gas value (unity) due to both the statistics of particles 
and the interaction.  

\section{ Concluding remarks}
\label{conclusion}

We have developed a phenomenological approach to 
a quantum liquid of particles obeying fractional exclusion 
 statistics.   
For generic local statistics of a fermionic  type, determined 
 by  (\ref{generic-distribution}), the starting basic 
equation for the distribution function of quasiparticles 
(\ref{basic-ES}) should be  replaced by 
\begin{equation}
\frac{ n_{\bbox{p} } } { {\cal F}(n_{\bbox{p}})  }=
e^{(\mu- {\varepsilon}_{\bbox{p}}[ n_{\bbox{p}}])/k_B T } \;.
\label{basic-general}\end{equation}
We note that in fact  all the formulas for the zero temperature quantities 
in Sections \ref{spinless} and 
\ref{spinning} as well as the formulas for the linear response 
(Sect.~\ref{linearResponse})
are valid in this more general   case 
provided $g$ is understood not as the exclusion 
statistics parameter but just as the height
of the step determining the distribution function of quasiparticles at 
zero temperature (\ref{n(T=0)}).    

The form of the transport equation (\ref{transport})
also holds for generic local statistics. 
We stress  that the Lorentz ratio---when it is determined by the  
scattering by impurities---only contains information about 
the statistics of quasiparticles thus 
being a good tool for search for  new statistics. 
In contrast,  the Wilson number encodes information both on interaction and 
statistics and cannot be interpreted in that simple way.  

We have discussed the usual, linear dispersion of quasiparticles 
near the pseudo-Fermi surface.  In fact, one can consider 
more complicated  dispersion laws. 
We comment in this respect on a possible application 
of ideas of generalized quantum liquids to models   
of two-dimensional fermions coupled to a  gauge field that were used to 
describe metallic non-Fermi liquid states.
  
Specifically, we consider a system of 
 fermions coupled to an abelian  Chern-Simons field 
and interacting with a two-body potential  
$v(\bbox{r}) \propto V_0 /r^{\eta}$ 
(or in Fourier components, 
 $V({\bbox q}) = V_0 / q^{2-\eta}$), 
with $1 \le \eta \le 2$ ($\eta =1 $ corresponds to the Coulomb interaction). 
This model is relevant to compressible quantum Hall states with 
filling factor $\nu=\frac12$ \cite{HLR} 
as well as  to high-$T_c$ superconductivity \cite{highT}. 

In an attempt to fit the result of the random phase approximation   
for the above model into the quasiparticle picture in the framework of the 
Landau Fermi liquid theory,   
a renormalized  energy of quasiparticles near the Fermi surface 
has been introduced, 
of the form     
\cite{HLR}
\begin{equation}
{\varepsilon}_{\bbox{p}}-\mu  \propto |p-p_{F}|^{\frac{1+\eta}{2}}\, 
{\rm sgn}\, (p-p_{F}) \, . 
\label{spectrum}\end{equation}
The associated  specific heat 
has the form  
$C= \gamma T^{\frac{2}{1+\eta}}$. 
However, the coefficient $\gamma$ differs 
from the result obtained in another way, 
from the free energy of the gauge field 
(the latter method, being gauge invariant, is regarded to be exact)
\cite{KimLee}. 
The reason for such inconsistency is in that the Landau quasiparticles 
in this system are ill-defined since the imaginary part of the self-energy of 
quasiparticles is of order of their real part (except  for $\eta =1$, where 
the damping of quasiparticles is logarithmically small compared to their 
energy, in which case the two ways to calculate the specific heat agree). 
 
For the quantum liquids discussed in the present paper, 
if  quasiparticles 
have the dispersion (\ref{spectrum}) (with $p_F \to p_0$),  
the temperature dependence of the specific heat will be  the same as for 
a Fermi liquid, with the coefficient  $\gamma$  
depending on the statistics of quasiparticles. 
Based on this, we point out a principal possibility  
to reconcile the results of 
calculations of the specific heat in the above system in  the 
two ways, by assuming that quasiparticles 
(for $\eta \neq 1$) form a non-Fermi 
quantum liquid, corresponding to some generalized statistics. 
In more general terms, our conjecture is that fluctuations of the gauge fields 
may lead to a kinematic effect (changing  statistics of excitations  
from Fermi to some other) for   metallic states.  
More detailed discussion of this issue is left for a future publication.

\bigskip 

\centerline{\bf Acknowledgments}

\medskip

I would like to thank D.P. Arovas for 
stimulating discussions and comments. 
Centre for Advanced Study (SHS) in Oslo 
is acknowledged 
for warm hospitality and support during the 1995/1996 program 
``Quantum phenomena in low-dimensional 
systems'' where this work was started. 
I also thank  
J.M. Leinaas and A.P. Polychronakos  for insightful discussions 
following 
my lectures on the Landau Fermi liquid theory at SHS.  
The work was partly  supported by the Norwegian Research Council.

\end{document}